\def\Ha{H$\alpha$}
\def\hii{\hbox{H{\thinspace}II}}
\def\hi{\hbox{H{\thinspace}I}}
\def\spose#1{\hbox to 0pt{#1\hss}}
\def\lta{\mathrel{\spose{\lower 3pt\hbox{$\mathchar"218$}}
     \raise 2.0pt\hbox{$\mathchar"13C$}}}
\def\gta{\mathrel{\spose{\lower 3pt\hbox{$\mathchar"218$}}
     \raise 2.0pt\hbox{$\mathchar"13E$}}}
\def\etal{{\it et\thinspace al.}~}
\def\eg{{\it e.g.,}~}

\documentstyle[11pt,paspconf]{article}

\begin{document}

\title{Superbubbles in the Magellanic Clouds}
\author{M. S. Oey}
\affil{Institute of Astronomy, Madingley Road, Cambridge CB3 0HA, U.K.}

\begin{abstract}
Superbubbles that result from the stellar winds and supernovae of
OB associations probably play a fundamental role in
the structure and energetics of the ISM in
star-forming galaxies.  Their influence may also dominate the
relationship between the different interstellar gas
phases.  How do superbubbles form and evolve?  How do they affect the
local and global ISM?  The Magellanic Clouds provide a superior
opportunity to study this shell-forming activity, since both
stellar content and gaseous structure can be examined in detail.
Here, the results of recent studies of superbubbles in the Magellanic
Clouds are reviewed. 
\end{abstract}


\section{Introduction}

It is now well-established that the kinetic feedback from massive
stars creates shell structures in the interstellar medium (ISM).
Indeed, studies of these objects 
in the Magellanic Clouds provide the best empirical understanding
of this phenomenon.  These types of shells can be roughly
classified into three categories (\eg Chu 1996; Meaburn 1980), which
are summarized in Table~1:  ``bubbles'' and supernova remnants (SNRs),
which result from stellar winds and supernovae (SNe) of individual massive 
stars; ``superbubbles,'' which result from the
action of a few, to hundreds of, OB star winds and SNe
clustered in an OB association; and ``supergiant shells,'' which 
have sizes of order $\sim 1$ kpc or more.  If the supergiant shells
are analogously created by massive star feedback, they must be
associated with starburst phenomena having total wind and SN energies
of $\gta 10^{54}$ erg.  This review will focus on recent (1990's)
studies of superbubbles in the Magellanic Clouds, with also a brief
look at supergiant shells. 

\begin{table}
\caption{OB star shells.} \label{tbl-1}
\begin{center}\scriptsize
\begin{tabular}{llcc}
Type & Parent population & $\log E$/erg & $\log R/$pc \\
\tableline
Bubble, SNR & single O, WR & 51 & 0 -- 1 \\
Superbubble & OB assoc.    & 52 -- 53 & 1 -- 2 \\
Supergiant shell & starburst & $\geq$ 54 & 2 -- 3 \\
\end{tabular}
\end{center}
\end{table}

How do these shells evolve?  Depending on the dominant physics, there
are three self-similar relations that are usually used to describe the
shell evolution.  In the case of a single point energy deposition, as
would apply to a SN, we have the traditional Sedov (1959) blastwave:
\begin{equation}\label{sedov}
R \propto (E/n)^{1/5}\ t^{2/5} \quad ,
\end{equation}
where $R$ is the shell radius, $E$ is the input mechanical energy, $n$
is a uniform ambient density, and $t$ is elapsed time.
For a constant mechanical power $L$ over an extended period, there is a
standard, adiabatic evolution (Pikel'ner 1968) given by,
\begin{equation}\label{ad}
R \propto (L/n)^{1/5}\ t^{3/5} \quad ,
\end{equation}
This model has
a double-shock structure, in which the inner shock, close to the
energy source, heats the wind ejecta to temperatures of order $10^6$ K,
driving the shell growth by the thermal pressure behind the outer shock at
the shell.  If the hot interior is able to radiate away its energy,
the inner shock will collapse upon the outer one, and the shell may
assume a momentum-conserving model (Steigman {\etal}1975):
\begin{equation}
R \propto (L/nv_\infty)^{1/4}\ t^{1/2} \quad ,
\end{equation}
where $v_\infty$ is the wind terminal velocity.  This model describes
a growth rate intermediate between those of equations~\ref{sedov}
and \ref{ad}.  There are other variations of these models; for example,
an extensive review of astrophysical blastwaves is given by Ostriker
\& McKee (1988). 

Assuming coeval star formation in OB associations,
the growth of superbubbles quickly becomes dominated by
SN activity, although most of those found in \Ha\ are
extremely young objects that are still wind-dominated, $\lta 5$
Myr old.

\section{Evolution}

Most empirical, recent studies of superbubbles attempt to test the
applicability of the standard, adiabatic model.  The LMC in
particular, offers a fine selection of objects, which have been used
to evaluate the shell dynamics and search for enclosed hot gas.

Recent dynamical studies (Oey \& Massey 1995; Oey 1996; Mac Low
{\it et al.} 1998; Oey \& Smedley 1998) of individual LMC superbubbles have
found a class of objects whose observed $v/R$ is far too high for
self-similar growth, in view of the stellar population present ($v$ is
the shell expansion velocity).  Based on excess observed X-ray
luminosities, Chu \& Mac Low (1990) and Wang \& Helfand (1991a) suggest
that most of these objects suffered recent SNR impacts on the
shell walls.  Complex velocity structures in
several objects (Rosado {\etal}1990; Ambrocio-Cruz {\etal}1997)
are consistent with this interpretation.  
However, Oey \& Smedley (1998) and Mac Low {\etal}(1998) demonstrate
that blowout from a high-density into a low-density region can
perfectly mimic the shell kinematics.

Although many optical superbubbles show this high $v/R$ evidence of
disruption, it is important to note that other objects have been found
whose dynamics are fully consistent with self-similar growth (Oey
1996).  However, all of the objects, along with the high $v/R$ objects, 
appear to indicate a growth rate discrepancy from the standard model,
where the assumed $L/n$ is overestimated by up to an order of
magnitude.  It is unclear whether this is caused by an overestimate in
stellar wind power, or underestimate in ambient $n$, or both.

The search for hot gas within the superbubbles is an important test of
the standard model.  Several studies (Chu \& Mac Low 1990; Wang \&
Helfand 1991a; Magnier {\etal}1996; Mac Low {\etal}1998) have found
objects with X-ray emission in excess of predictions.  As mentioned
above, these may be caused by SNR impacts.  Reassuringly, there are other
superbubbles that do not show X-ray enhancements (Chu {\etal}1995),
and detection limits for these remain consistent with the predicted
X-ray emission.  Furthermore, in all superbubbles that have been
examined for C IV and Si IV absorption, these high-ionization species
have been found (Chu {\etal}1994), and can be attributed to an
interface layer between the shell wall and hot interior.

\section{Supergiant Shells}

Supergiant shells (SGS's) are more poorly understood than
superbubbles.  It is possible that many of these do not originate from
star formation events at all.  Alternative mechanisms include impacts by
high-velocity clouds (\eg Tenorio-Tagle {\etal}1987) and gamma-ray bursters
(Efremov {\etal}1998; Loeb \& Perna 1998).  At any rate, the evolution of SGS's
is likely to differ substantially from that of superbubbles
because the size scales are similar to those of galactic
parameters like scale height and length.

The SGS's in the LMC have been catalogued by Meaburn (1980), and only
two of these have been examined in any detail.  The most well-known
example is LMC-4, associated with the Constellation III stellar
region.  This region is a remarkably well-defined shell,
$\sim 600$ pc in radius, seen in both \hi\ and as a ring of optical
\hii\ regions.  There have been numerous studies of LMC-4 over the
last two decades; the most recent work is by Domgoergen {\etal}(1995)
and Bomans {\etal}(1996) studying its gaseous properties, and by Olsen
{\etal}(1997) and Braun {\etal}(1997) on the stellar content.
The other LMC SGS that has been examined is LMC-2, just east of 30
Dor.  Its gas properties were studied by Wang \& Helfand (1991b) and
Caulet \& Newell (1996).

\section{Superbubbles and the Global ISM}

Superbubbles, as a direct manifestation of massive star feedback,
are of critical importance to the global properties of the ISM.  These
include ISM structure and kinematics, relationship between the
different gas phases, and interstellar processes like star formation.

To gain insight on the ISM structure and kinematics caused by
superbubbles, it is possible to use equation~\ref{ad} to derive a size
and velocity distribution for superbubbles in the ISM.  This also
serves as an additional test of the standard, adiabatic evolution over the
full lifetimes of the superbubbles.  Oey \& Clarke (1997) computed the
size distribution $N(R)\thinspace dR\propto R^{-\alpha} dR$ for 
the SMC, using the \hii\ region luminosity function and assuming a
constant creation rate for OB associations.  The observed slope
$\alpha=2.7\pm 0.6$ for \hi\ superbubbles (Staveley-Smith {\etal}1997)
is in remarkable agreement with the predicted slope of $2.8\pm 0.4$,
suggesting that superbubble activity can fully explain the \hi\
structure of the SMC.  This can be compared to $N(R)\thinspace dR$ for holes in
a fractal ISM structure (Elmegreen 1997), which is being quantified
for the SMC by Stanimirovi\'c {\etal}(1998).  They find a volume fractal 
dimension $D=2.5$, translating directly to $\alpha=2.5$.  It is
therefore too difficult to distinguish between a superbubble and
fractal ISM model, using only $N(R)\ dR$.  Indeed, the similarity in
observed $N(R)\ dR$ for both cases in the SMC, might be indicative of
a physical relation between the two structural models.

It is also possible to test for superbubble structure in the ISM by
examining the morphology of the \hi\ holes, of which $>500$ were catalogued
as shells, based on elliptical shape (Staveley-Smith
{\etal}1997).  In addition, the velocity distribution
$N(v)\thinspace dv$ computed by Oey \& Clarke (1998) is also
consistent with the observed shell velocities for the SMC.  
A superbubble velocity structure could be a dominant input to
the kinematics and turbulence of the ISM.

As superbubbles expand and age, their surface brightness quickly
diminishes.  Hence, they could be an important component of the
diffuse, warm, ionized medium (WIM).  Hunter (1994) obtained
emission-line spectra of superbubbles in the LMC, which do
show that some of the line ratios are intermediate between those for
classical \hii\ regions and WIM detected in other galaxies.  Kennicutt
{\etal}(1995) examined the morphological structure of the WIM in the
Magellanic Clouds, and also found a substantial contribution from
faint superbubbles and SGS's. 

Finally, it is widely thought that superbubbles can trigger renewed star
formation in their shells of swept-up cool gas.  Several candidates
have been examined in the LMC.  The clearest example is DEM~34 / N11
(Walborn \& Parker 1992; Rosado {\etal}1996), and additional
candidates are DEM~152 / N44 (Oey \& Massey 1995) and DEM~192 / N51~D
(Oey \& Smedley 1998).  As mentioned above, the SGS LMC-4 has a long
history of investigation as a triggered star formation candidate (\eg
Efremov 1998). 

\acknowledgments

I am grateful for an IAU Travel Grant that made possible my participation
in this meeting.


\begin{references}
\reference Ambrocio-Cruz, P., Laval, A., Marcelin, M., \& Amram,
	P. 1997, \aap, 319, 973
\reference Bomans, D. J., De Boer, K. S., Koornneef, J., \& Grebel,
	E. K. 1996, \aap, 313, 101
\reference Braun, J. M., Bomans, D. J., Will, J.-M., \& De Boer,
	K. S. 1997, \aap, 328, 167
\reference Caulet, A. \& Newell, R. 1996, \apj, 465, 205
\reference Chu, Y.-H. 1996, in {\sl The Interplay between Massive Star
	Formation, the ISM, and Galaxy Evolution,} eds. D. Kunth,
	B. Guiderdoni, M. Heydari-Malayeri, \& T. X. Thuan,
	Gif-sur-Yvette:  Editions Fronti\`eres, 201
\reference Chu, Y.-H. \& Mac Low, M.-M. 1990, \apj, 365, 510
\reference Chu, Y.-H., Chang, H.-W., Su, Y.-L., \& Mac Low,
	M.-M. 1995, \apj, 450, 157
\reference Chu, Y.-H., Wakker, B., Mac Low, M.-M., Garc\'\i a-Segura,
	G. 1994, \aj, 108, 1696
\reference Domgoergen, H., Bomans, D. J., \& De Boer, K. S. 1995,
	\aap, 296, 523
\reference Efremov, Y. N. 1998, this volume
\reference Efremov, Y. N., Elmegreen, B. G., \& Hodge, P. W. 1998,
	\apjl, 501, L163
\reference Elmegreen, B. G. 1997, \apj, 477, 196
\reference Hunter, D. A. 1994, \aj, 107, 565
\reference Kennicutt, R. C., Bresolin, F., Bomans, D. J., Bothun,
	G. D., \& Thompson, I. B. 1995, \aj, 109, 594
\reference Loeb, A. \& Perna, R. 1998, \apjl, 503, L35
\reference Mac Low, M.-M., Chang, T. H., Chu, Y.-H., Points, S. D.,
	Smith, R. C., \& Wakker, B. P. 1998, \apj, 493, 260
\reference Magnier, E. A., Chu, Y.-H., Points, S. D., Hwang, U., \& 
	Smith, R. C. 1996, \apj, 464, 829
\reference Meaburn, J. 1980, \mnras, 192, 365
\reference Oey, M. S. 1996, \apj, 467, 666
\reference Oey, M. S. \& Clarke, C. J. 1997, \mnras, 289, 570
\reference Oey, M. S. \& Clarke, C. J. 1998, in {\sl Interstellar
	Turbulence,} eds. J. Franco \& A. Carraminana, Cambridge:
	Cambridge Univ. Press, in press
\reference Oey, M. S. \& Massey, P. 1995, \apj, 452, 210
\reference Oey, M. S. \& Smedley, S. A. 1998, \aj, in press
\reference Olsen, K. A. G., Hodge, P. W., Wilcots, E. M., \& Pastwick,
	L. 1997, \apj, 475, 545
\reference Ostriker, J. P. \& McKee, C. F. 1988, Rev. Mod. Phys, 60, 1
\reference Pikel'ner, S. B. 1968, Astrophys. Lett., 2, 97
\reference Rosado, M. {\etal}1990, \aap, 238, 315
\reference Rosado, M. {\etal}1996, \aap, 308, 588
\reference Sedov, L. I. 1959, {\sl Similarity and Dimensional Methods
	in Mechanics}, New York:  Academic
\reference Stanimirovi\'c, S., Staveley-Smith, L., Dickey, J. M.,
	Sault, R. J., \& Snowden, S. L. 1998, this volume
\reference Staveley-Smith, L., Sault, R. J., Hatzidimitriou, D.,
Kesteven, M., \& McConnell, D. 1997, \mnras, 289, 225
\reference Steigman, G., Strittmatter, P. A., \& Williams, R. E. 1975,
	\apj, 198, 575
\reference Tenorio-Tagle, G., Franco, J., Bodenheimer, P., \&
	R\'o\.zyczka, M. 1987, \aap, 179, 219
\reference Walborn, N. R. \& Parker, J. W. 1992, \apjl, 399, L87
\reference Wang, Q. \& Helfand, D. J. 1991a, \apj, 373, 497
\reference Wang, Q. \& Helfand, D. J. 1991b, \apj, 379, 327
\end{references}
\end{document}